\documentclass[a4paper,11pt]{article} 
\usepackage{jheppub}

\usepackage[utf8]{inputenc}
\pdfoutput=1 
\usepackage[T1]{fontenc} 
\usepackage{cancel}
\usepackage{bm}
\usepackage{comment} 
\usepackage{braket}

\begin{document}
\title{String Excitation by Initial Singularity of Inflation}

\author{Kanji Nishii,}
\emailAdd{kanji.nishii@stu.kobe-u.ac.jp} 
\affiliation{Department of Physics, Kobe University, Kobe 657-8501, Japan}

\author{Daisuke Yoshida}
\emailAdd{dyoshida@hawk.kobe-u.ac.jp}

\abstract{
We discuss excitation of string oscillation modes by an initial singularity of inflation. The initial singularity of inflation is known to occur with a finite Hubble parameter, which is generally lower than the string scale, and hence it is not clear that stringy effects become significant around it. With the help of Penrose limit, we find that infinitely heavy oscillation modes get excited when a singularity is strong in the sense of Krolak's classification. We demonstrate that the initial singularities of Starobinsky and hill top inflation, assuming the slow roll inflation to the past infinity, are strong. Hence stringy corrections are inevitable in the very early stage of these inflation models. We also find that the initial singularity of the hill top inflation could be weak for non-slow roll case. 
} 

\preprint{KOBE-COSMO-21-09}

\maketitle
\flushbottom 

\section{Introduction}
The presence of an inflationary stage of the Universe is strongly supported by the cosmic microwave background observations~\cite{Akrami:2018odb}.  The observation can access the period of the last 60 e-folding in inflationary era and we hardly know about a stage of universe before, or in the very beginning of the inflation. From the theoretical view point, the Universe must be in another phase before the inflation. In Ref.~\cite{Borde:2001nh}, it is shown that the spacetime region that is covered by the world lines of comoving observers who feel accelerated expansion is past incomplete. The past incompleteness suggests the presence of a boundary for inflationary spacetime. Then if the boundary is inextendible, the inflation spacetime is said to have an initial singularity, otherwise it is a coordinate singularity.

Since a singularity is expected to be a sign of violation of classical gravity, it is important to ask whether the past boundary of an inflationary universe is a singularity or not. In Ref.~\cite{Yoshida:2018ndv}, the past extendibility of accelerated flat Friedmann--Lema\^itre--Robertson--Walker (FLRW) universes is investigated by checking the presence of a parallelly propagated curvature singularity~\cite{Hawking:1973uf, Ellis:1977pj}, which is defined by the divergence of the components of the Riemann tensor with respect to the basis which is parallel transportation along the incomplete geodesics. Then it was found that the extendibility depends on whether $\lim_{t \rightarrow -\infty} \dot{H}/a^2 $ is finite or not, where $a$ is the scale factor, $H$ is the Hubble parameter and $t$ is the proper time of the comoving observers. By applying this method, for example, the Starobinsky's inflation model is found to have an initial parallelly propagated curvature singularity assuming inflation persists on the slow roll trajectory toward $t \rightarrow - \infty$. 
See also Refs.~\cite{FernandezJambrina:2007sx, Fernandez-Jambrina:2016clh} where the initial singularity of other type inflation like a power law was studied. The analysis of extendibility in Ref.~\cite{Yoshida:2018ndv} has been then extended to the universe with a spatial curvature or anisotropies recently in Ref.~\cite{Nomura:2021lzz}.

Interestingly, the singularity of inflationary universe is quite different from the big-bang singularity in the sense that there is no divergence with the curvature invariants. Thus this singularity occurs even when the typical energy scale read from the Hubble parameter is much smaller than the string scale. Therefore it is not clear that this kind of singularity leads the violation of low energy effective theory description of inflation model. The purpose of this paper is to study the propagation of strings on inflationary universe to answer this question.

The equations of motion for string in de Sitter spacetime was solved exactly by Refs.~\cite{Combes:1993rw, deVega:1993rm}. Recently, this analysis is revisited to see the consistency between the Higuchi bound and the modified Regge trajectory \cite{Noumi:2019ohm, Lust:2019lmq, Kato:2021rdz, Parmentier:2021nwz}. String propagation in an expanding universe has been studied in many contexts so far \cite{Sanchez:1989cw,Gasperini:1990xg,Larsen:1995vr,deVega:1995bq, Tolley:2005us}. Contrary to the exact de Sitter case, one need to take some approximation to study the string on general expanding universe. In this paper, we will focus on the method of Penrose limit \cite{Penrose1976}, as did in \cite{Tolley:2005us}.

The Penrose limit is a way to obtain a plane wave spacetime spreading an infinitesimal neighborhood of a null geodesic outward. It has been studied actively as a way to construct nontrivial supersymmetric solutions \cite{Blau:2002dy} from known supersymmetric spacetimes. This is because the Penrose limit does not decrease a number of symmetry \cite{Geroch:1969ca, Blau:2002mw}. A class of spacetime with a singularity is expected to have a kind of universality~\cite{Blau:2003dz,Blau:2004yi}. As we will see later, an initial singularity of an inflationary universe is out of this universality. 

String propagation in a plane wave spacetime with a singularity has actively studied so far~\cite{Horowitz:1989bv,deVega:1990kk,Horowitz:1990sr,deVega:1990ke,David:2003vn,Craps:2008bv}. An important feature found in Ref. \cite{David:2003vn} is that the quantum string propagation is possibly ill-defined even when a singularity is weak in the sense of Tipler's classification. Actually, we will see later, that consistency of string propagation can be judged by Krolak's classification, not the Tipler's one. Note that the singular behavior becomes milder for a point-like shock wave spacetime \cite{Aichelburg:1970dh, Veneziano:1987df, Amati:1988ww, deVega:1988ts, deVega:1990nr, Sanchez:2003ek}, not a plane wave.

This paper is organized as follows.
In Sec.~\ref{sec2}, we examine the presence/absence of an initial singularity of inflationary Universe based on Ref.~\cite{Yoshida:2018ndv} and discuss its past completion and the Penrose limit. As a result of Penrose limit, singular/non-singular inflationary Universe respectively reduces to singular/non-singular plane wave spacetime.  In the Sec.~\ref{sec3}, the propagation of test string in a plane wave spacetime is discussed. There we find that string can propagate through the initial singularity of the Universe if the strength of the singularity is sufficiently weak.  In the Sec.~\ref{sec4}, we study the strength of the singularity for three models of inflation; Starobinsky inflation and general and quadratic hill top inflation. The final section is devoted to summary and discussion.

\section{Inflationary Universe and its past completion}
\label{sec2}
Here, we review the presence/absence of an initial singularity of inflationary Universe in Sec.~\ref{sec21}. Then we construct a past completion in Sec.~\ref{sec22} and take the Penrose limit  to obtain singular/non-singular plane wave spacetime in Sec.~\ref{sec23}.

\subsection{Initial singularity of inflation}
\label{sec21}
We consider $3+1$ dimensional flat FLRW universe, 
\begin{align}
 g_{\mu\nu} d x^{\mu} d x^{\nu} &= - d t^2 + a(t)^2 (d r^2 + r^2 d \Omega^2) = a(\eta)^2 \left( - d \eta^2 + dr^2 + r^2 d \Omega^2 \right), \label{FLRW}
\end{align}
with $d \Omega^2 := d \theta^2+\sin^2\theta d \phi^2$.
Here $a$ is the scale factor and $t$ represents the comoving time while $\eta$ represents the conformal time, $d t = a d \eta$.

We assume that the scale factor $a(t)$ approaches to zero toward a initial time $t_{i}$: $\lim_{t \rightarrow t_{i}} a(t) = 0$. To discuss the (in)completeness and extendibility of null geodesics at $a \rightarrow 0$,  we consider null geodesics defined by $\eta + r = \text{constant}$. The affine parameter $u$ of such geodesics can be defined by
\begin{align}
 du = a^2 d\eta = a dt.
\end{align}
Thus the incompleteness can be checked by
\begin{align}
 \int^{t_{i}} a(t) dt = \text{finite.} \Leftrightarrow  \text{null incomplete}.
\end{align}
In the following, we focus on the incomplete case and fix the integration constant by $u(t_{i}) = 0$. The incompleteness of the null geodesics implies two possibilities. One is that the whole of the spacetime is not spanned by the flat FLRW coordinates \eqref{FLRW} and the end point of null geodesics corresponds to a coordinate singularity. A simple example of the extendible case is the exact de Sitter space, where flat chart only covers a half of the entire closed de Sitter space. The other possibility is that the end point of null geodesics corresponds to a true singularity.

The tangent vector $k^\mu =dx^\mu/du$ of such null geodesics is given by
\begin{align}
 k^{\mu} \partial_{\mu}
= \frac{1}{a^2} \partial_{\eta} - \frac{1}{a^2} \partial_{r}
 = \frac{1}{a} \partial_{t} - \frac{1}{a^2} \partial_{r}.
\end{align}
Because of the relation $\nabla_{[\mu} k_{\nu]} = 0$, we can introduce a scalar potential for $k$ by $k_{\mu} dx^{\mu} = - d v $, that is, we introduce $v$ by
\begin{align}
 dv := - k_{\mu} dx^{\mu}  =  \frac{1}{a} dt + dr = d\eta + dr.
\end{align}
Following, we fix an integration constant by $v = \eta + r$. Note that a curve $v = \text{constant}$ corresponds to a null geodesic.

By using $u$ and $v$ coordinates instead of $t$ and $r$, we can rewrite \eqref{FLRW} as
\begin{align}
 g_{\mu\nu}dx^{\mu} dx^{\nu} &= - 2 du dv + a^2(u) dv^2 + a^2(u) (v - \eta(u))^2 d \Omega^2.
\label{g=}
\end{align}
In this coordinates, the Ricci tensor can be represented as
\begin{align}
 R_{\mu\nu} dx^{\mu} dx^{\nu} &= - 2 \frac{a''(u)}{a(u)} du^2 + \left(3 a'(u){}^2 + a(u) a''(u)\right)g_{\mu\nu}dx^{\mu} dx^{\nu} \notag\\
&= - 2 \frac{\dot{H}}{a^2}  du^2 + \left( 3 H^2 + \dot{H} \right) g_{\mu\nu}dx^{\mu} dx^{\nu},
\end{align}
where the prime stands for derivatives with respect to $u$ and $H$ is the Hubble parameter defined by $H := \dot{a}/ a = a'$.
The expression appeared in $R_{uu}$ plays an important role in this paper and let us call it $A(u)$: 
\begin{align}
 A(u) :=- \frac{1}{2} R_{uu}=\frac{a''(u)}{a(u)} = \frac{\dot{H}}{a^2}. \label{Au}
\end{align}

One can see that the null tetrad basis
\begin{align}
e^{u} := 
du - \frac{1}{2} a^2 dv,  \qquad 
e^{v} := 
dv, \qquad
e^{\theta} := 
d \theta, \qquad
e^{\phi} := 
\sin \theta d\phi,
\end{align}
are parallelly transportation along null geodesics. By the definition, the divergence of the components of the Riemann tensor with respect to parallelly propagated basis corresponds to a parallelly propagated curvature singularity \cite{Ellis:1977pj,Hawking:1973uf}.
Since $e^{u} \simeq du $ in the limit $a \rightarrow 0$, we obtain,
\begin{align}
 R_{\mu\nu} dx^{\mu} dx^{\nu} \overset{a \rightarrow 0}{\simeq} - 2 A(u) e^{u} e^{u} + \left(3 H^2 + \dot{H} \right) \eta_{\mu\nu} e^{\mu} e^{\nu}. \label{Ricci}
\end{align} 
Thus $a \rightarrow 0$ corresponds to a parallelly propagated curvature singularity if $A(u)$ or $3 H^2 + \dot{H} $ diverges.

\begin{figure*}
\centering
 \includegraphics[width=0.5\hsize]{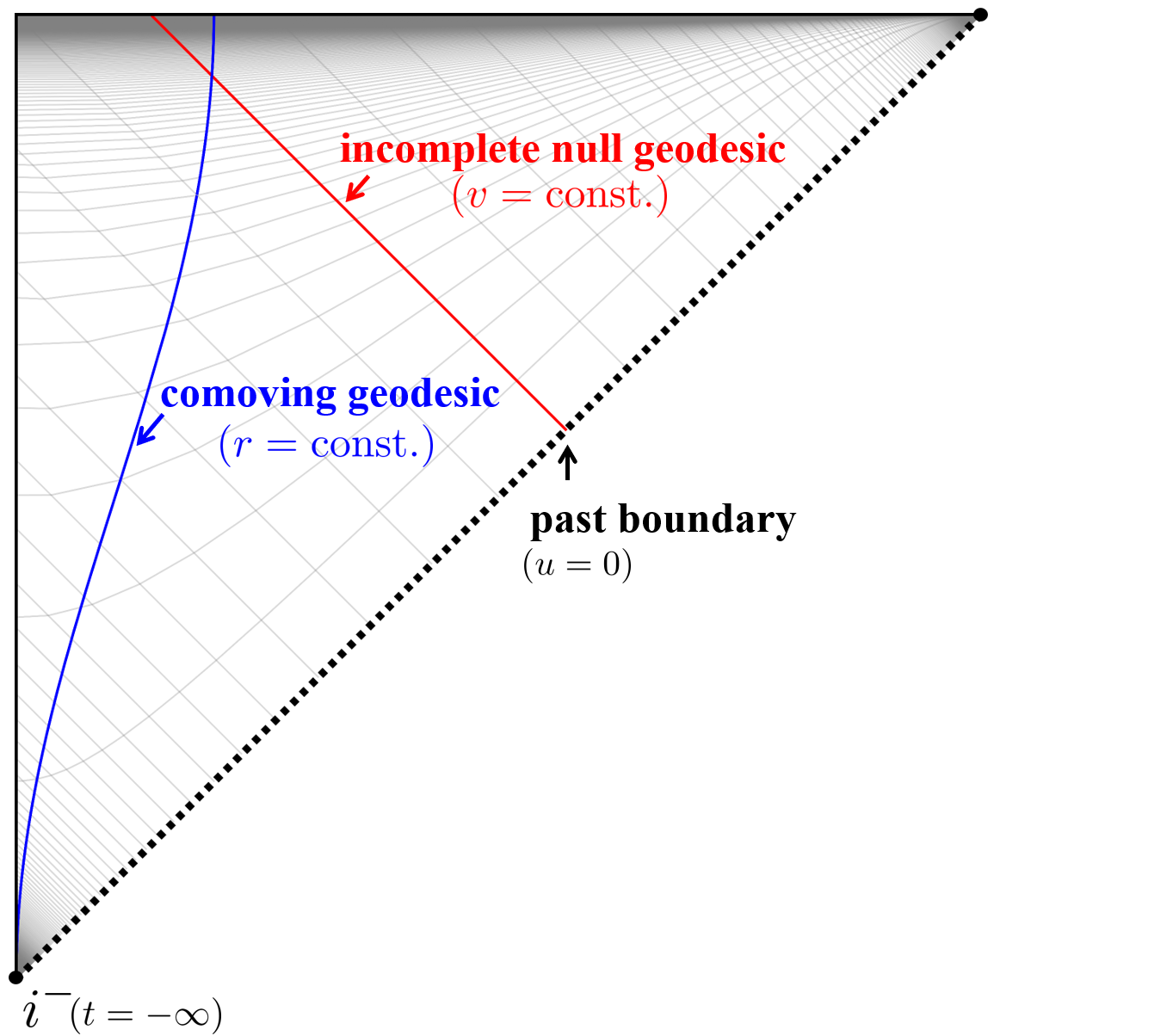}
 \caption{A typical Penrose diagram of quasi de Sitter spacetime:
The comoving geodesic ($r = \text{constant}$) is past complete and approaches $i^{-}$ in an infinite proper time $t$.
On the other hand, the null geodesic ($v = \text{constant}$) is past incomplete and approaches the past boundary (the dashed line) in a finite affine length $u$. The metric has at least continuous extension beyond the past boundary, though the curvature could be a parallelly propagated curvature singularity depending on the scale factor. We note that this is a diagram with future complete null geodesics, though the future structure is irrelevant to our discussion in this paper. 
}
\label{fig1}
\end{figure*}

In the following, we focus on an inflationary universe which approaches flat de Sitter Universe in the limit $t \rightarrow - \infty$,
\begin{align}
 a(t) \simeq \mathrm{e}^{\bar{H} t} + \cdots.
\label{pastdS}
\end{align}
This scale factor satisfies the assumptions which we have assumed above: vanishing scale factor at $t \rightarrow t_{i} := - \infty$ and the incompleteness of the null geodesics. The timelike comoving geodesics are past complete because $t$ is the proper time of these geodesics. A typical Penrose diagram of this kind of universe is shown in Fig.~\ref{fig1}. 

The assumption \eqref{pastdS} means that $H$ and $\dot{H}$ are finite as $ a \rightarrow 0$. Thus only the dangerous component in Eq.~\eqref{Ricci} is the one with $A(u)$.  
Therefore, with the assumption \eqref{pastdS}, the past boundary (the dashed line in Fig.~\ref{fig1}) is a parallelly propagated curvature singularity if and only if 
\begin{align}
 \lim_{t \rightarrow - \infty} \frac{\dot{H}}{a^2}= \lim_{u \rightarrow 0} A(u) = \pm \infty. \label{limA}
\end{align} 
The limit \eqref{limA} depends on the next to leading order of the scale factor omitted in Eq.~\eqref{pastdS} and hence depends on the detail of inflationary model.
For example, $A(u) = 0$ for the exact flat de Sitter space $a \propto \mathrm{e}^{\bar{H} t}$ and hence the past boundary is extendible as expected. 

An interesting property of this Universe is that the metric components are well defined in the limit $u \rightarrow 0$,
\begin{align}
 g_{\mu\nu} dx^{\mu} dx^{\nu} \overset{u \rightarrow 0}{\sim} - 2 du dv  + \frac{1}{\bar{H}^2} d\Omega^2 ,
\end{align}
as one can evaluate
\begin{align}
 \lim_{u \rightarrow 0} a(u) \eta(u) =  - \frac{1}{\bar{H}}.
\end{align}
Thus the metric has continuous extension beyond the boundary, even though it may not be smooth. We can define the further past of inflationary Universe beyond an ``initial'' singularity. 

\subsection*{An example}
Let us consider an explicit example of an universe with the assumption \eqref{pastdS}, which can be understood as
\begin{align}
H = a'(u) \overset{u \rightarrow 0}{\sim} \bar{H} + \cdots.
\end{align}
Thus we obtain
\begin{align}
 a(u) = \bar{H} u + \cdots. \label{scaleu}
\end{align}
As pointed above, the property of the boundary is sensitive to the next to leading term. 
Here let us assume that the next to leading order term is power of $u$,
\begin{align}
a(u) = \bar{H} u \left( 1  - \frac{\kappa}{(3 - \beta)(2 - \beta)} u^{2 - \beta} + \cdots \right), \label{a(u)=}
\end{align}
with constants $\beta$ and $\kappa$. In order for this term to be sub-leading, $\beta$ is assumed to satisfy $\beta < 2$.
We will see in Sec.~\ref{sec4} that this scale factor corresponds to that of the quadratic hill top inflation.
Note that this scale factor can be expressed by the comoving time $t$ as
\begin{align}
 a(t) = \mathrm{e}^{\bar{H} t} - \frac{\kappa \bar{H}^{\beta - 2}}{(2- \beta)^2}  \mathrm{e}^{(3 - \beta) \bar{H} t} + \cdots. \label{a(t)=}
\end{align}

From the expression \eqref{a(u)=}, the curvature component $A(u)$ can be evaluated as
\begin{align}
 A(u) = \frac{a''(u)}{a(u)} = - \frac{\kappa}{u^{\beta} } + \cdots . \label{A(u)=}
\end{align}
Thus for a positive $\beta$, the boundary $u \rightarrow 0$ becomes a parallelly propagated curvature singularity while it is a regular boundary for a non-positive $\beta$. Since the sign of $A$ coincides with that of $\dot{H}$, a positive $\kappa$ corresponds to the expansion of the universe with the matter contents satisfy the null energy condition.

Let us check the implication of the parameter $\beta$ to the strength of the singularity formulated by Tipler and Krolak \cite{Tipler:1977zza, krolak1983proof, CLARKE1985127,Krolak_1986}. The strong/weak singularity in the sense of Tipler is defined by the divergence/convergence of the limit
\begin{align}
 \int^{0} du \int^{u} du' R_{uu}(u') \propto \lim_{u \rightarrow 0}  \frac{1}{u^{\beta - 2}},
\end{align}
while that in the sense of Krolak is defined with
\begin{align}
 \int^{0} du R_{uu}(u) \propto \lim_{u \rightarrow 0}  \frac{1}{u^{\beta - 1}}.
\end{align}
The classification of the strength of the singularity can be summarized in Table \ref{table1}.
\begin{table}[htb]
\begin{center}
\begin{tabular}{l||cllllllll}
$\beta$ &                             & \multicolumn{2}{c}{0}                        & \multicolumn{1}{c}{} & \multicolumn{2}{c}{1}                        &             & \multicolumn{2}{c}{2}    \\ \hline
p.p. curvature singularity   & \multicolumn{1}{l}{regular} & \multicolumn{1}{l|}{} & \multicolumn{1}{c}{} & singularity          & \multicolumn{1}{l|}{} &                      & singularity & \multicolumn{1}{l|}{} &  \\ \cline{1-1}
Krolak's classification  & regular                     & \multicolumn{1}{l|}{} &                      & weak                 & \multicolumn{1}{l|}{} & \multicolumn{1}{c}{} & strong      & \multicolumn{1}{l|}{} &  \\ \cline{1-1}
Tipler's classification  & regular                     & \multicolumn{1}{l|}{} &                      & weak                 & \multicolumn{1}{l|}{} &                      & weak        & \multicolumn{1}{l|}{} & 
\end{tabular}
\caption{classification of singularity}
\label{table1}
\end{center}
\end{table}

With the expression of the conformal time, 
\begin{align}
 \eta(u) = - \frac{1}{\bar{H}^2 u} \left(
1 + \frac{2 \kappa}{(\beta -1 )(\beta-2)(\beta-3)} u^{2 - \beta} + \cdots 
\right),
\end{align}
one can express the metric~\eqref{g=} near $u = 0$ as
\begin{align}
 g_{\mu\nu}dx^{\mu} dx^{\nu} 
&=
- 2 du dv + \bar{H}^2 u^2 \left(1 - \frac{2 \kappa}{(3 - \beta)(2 - \beta)} u^{2 - \beta} + \cdots \right) dv^2 \notag\\
& \qquad + \frac{1}{\bar{H}^2}  \left( \left( 1 + \bar{H}^2 u v \right)^2 - \frac{2 \alpha}{(2 - \beta)(1 - \beta)} u^{2 - \beta} + \cdots \right) d \Omega^2.
\end{align}
The metric components are well defined in the limit $u \rightarrow 0$ as expected.

\subsection{Contracting Universe as a past completion}
\label{sec22}
So far, we have seen that the metric components have at least continuous extension beyond the past boundary of the quasi de Sitter universe. In this section we provide an extension of the universe beyond the past boundary.

Let us consider a contracting FLRW universe which has a scale $\tilde{a}(\tilde{t})$ with FLRW coordinates $(\tilde{t}, \tilde{r}, \theta, \phi)$. The metric is given by
\begin{align}\label{contmetric}
 g_{\mu\nu}dx^{\mu} dx^{\nu} = - d\tilde{t}^2 + \tilde{a}(\tilde{t})^2 (d\tilde{r}^2 + \tilde{r}^2 d \Omega^2 )
\end{align}
and we assume the scale factor satisfies
\begin{align}
 \tilde{a}(\tilde{t}) = a( - \tilde{t}),
\end{align}
where $a$ is the scale factor $\tilde{a}(\tilde{t})$ of the expanding Universe. Similar to the case of expanding Universe,
we introduce the conformal time $\tilde{\eta}$ by $d\tilde{t} = \tilde{a} d \tilde{\eta}$, the affine length of a future out-going null geodesic by $\tilde{u}$ and the coordinate $\tilde{v}$ by $ \tilde{v} = \tilde{\eta} - \tilde{r}$. 
Then the metric \eqref{contmetric} can be written as
\begin{align}
 g_{\mu\nu}dx^{\mu} dx^{\nu} &= -2 d\tilde{u} d\tilde{v} + \tilde{a}^2(\tilde{u}) d \tilde{v}^2 + \tilde{a}^2(\tilde{u}) \left( \tilde{\eta}(\tilde{u}) - \tilde{v} \right)^2 d \Omega^2 \notag\\
& = -2 d\tilde{u} d\tilde{v} + a^2(- \tilde{u}) d \tilde{v}^2 + a^2(- \tilde{u}) \left( - \eta( - \tilde{u}) - \tilde{v} \right)^2 d \Omega^2 
\end{align}
where we used the relation, $ \tilde{a}(\tilde{u}) = a( - \tilde{u}), \tilde{\eta}(\tilde{u}) = - \eta(- \tilde{u}) $.

\begin{figure*}
\centering
 \includegraphics[width=0.5\hsize]{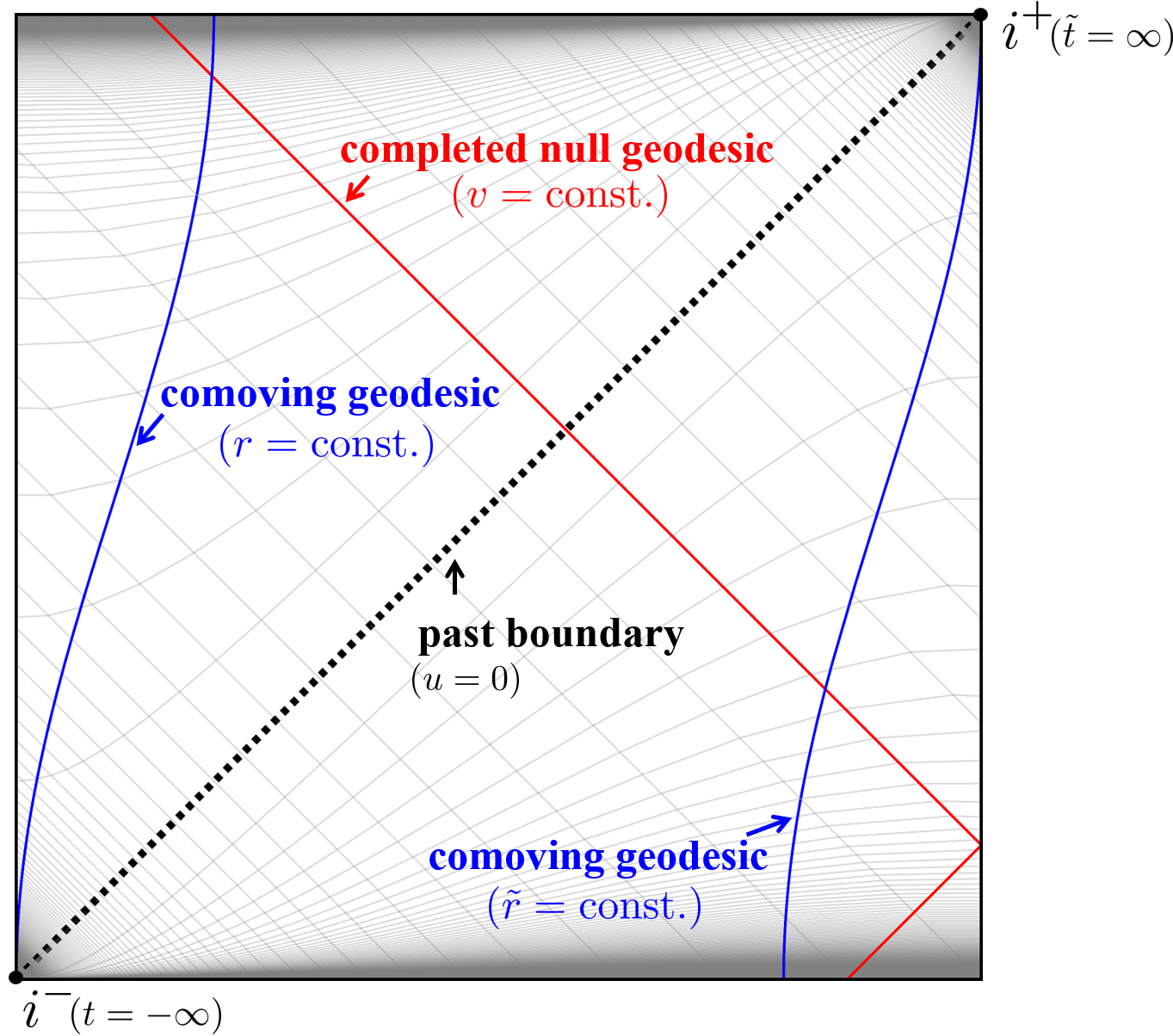}
 \caption{Penrose diagram of past completed universe:
The region above the boundary (dashed line) represents an expanding universe and that below the boundary represents a contracting universe. The comoving geodesics in the expanding universe ($r = \text{constant}$) does not enter the contracting region, contrary to the standard bouncing universe where comoving observers pass through both the expanding and contracting epoch. 
The null geodesic ($v = \text{constant}$) pass through the boundary of the original inflationary universe, where it could be a parallelly propagated curvature singularity depending on the model.}
\label{fig2}
\end{figure*}

Based on these coordinates, we consider a contracting Universe as a past completion of the inflationary Universe, by identifying $u < 0$ as $\tilde{u}$ and $v$ as $\tilde{v}$.
Thus the metric of the entire spacetime can be written as
\begin{align}
  g_{\mu\nu}dx^{\mu} dx^{\nu} &=
 \begin{cases}
  - 2 du dv + a^2(u) dv^2 + a^2(u) (\eta(u) - v)^2 d \Omega^2 &(u \geq 0) \\
  - 2 du dv + a^2(-u) dv^2 + a^2( - u) (- \eta(-u) - v)^2 d \Omega^2 &(u<0)
 \end{cases} 
\notag\\
&= - 2 du dv + a^2(|u|) dv^2 + a^2(|u|) (\text{sgn}(u) \eta(|u|) - v)^2 d \Omega^2.
\label{completedgeneralFLRW}
\end{align}
A typical Penrose diagram of this universe is given in Fig.~\ref{fig2}.

The Ricci tensor in whole of the Universe can be evaluated as
\begin{align}
R_{\mu\nu} dx^{\mu} dx^{\nu} &= - 2 \frac{a''(|u|)}{a(|u|)} du^2 + \left(3 a'(|u|){}^2 + a(|u|) a''(|u|)\right)g_{\mu\nu}dx^{\mu} dx^{\nu} .
\end{align}
Then we can describe a component of Ricci tensor in the whole of spacetime 
with generalizing the definition of $A(u)$ as
\begin{align}
 A(u < 0) = A(|u|).
\end{align} 
\subsection*{An example}
Again, let us focus on the example \eqref{a(u)=}. In this case the scale factor of the contracting universe is given by
\begin{align}
 \tilde{a}(\tilde{t}) = a(- \tilde{t}) =  \mathrm{e}^{- \bar{H} \tilde{t}} - \frac{\kappa \bar{H}^{\beta - 2}}{(2 - \beta)^2} \mathrm{e}^{- (3 - \beta) \bar{H} \tilde{t}} + \cdots \qquad (\tilde{t} \sim + \infty).
\end{align}
Now, we rewrite $\tilde{a}$ and $\tilde{\eta}$ as a function of $\tilde{u}$:
\begin{align}
 \tilde{a}(\tilde{u}) &= - \bar{H} \tilde{u} \Bigl(1  - \frac{\kappa }{(\beta - 3) ( \beta -2)} (- \tilde{u})^{2 - \beta} + \cdots \Bigr) = a( - \tilde{u}), \\
 \tilde{\eta}(\tilde{u}) &= - \frac{1}{\bar{H}^2 \tilde{u}} \left( 1 + \frac{2 \kappa}{(\beta - 1)(\beta - 2)(\beta - 3)} (- \tilde{u})^{2 - \beta} + \cdots \right) = - \eta( -\tilde{u}),
\end{align}
By introducing $\{ \tilde{u}, \tilde{v} \}$ coordinates, the metric of the contracting universe can be written as
\begin{align}
g_{\mu\nu}dx^{\mu} dx^{\nu} & =  - 2 d \tilde{u} d \tilde{v} +  \bar{H}^2 \tilde{u}^2 \Bigl(1  - \frac{2 \kappa }{(\beta - 3) ( \beta -2)} (- \tilde{u})^{2 - \beta} + \cdots \Bigr) d\tilde{v}^2 \notag\\
& \qquad  + \frac{1}{\bar{H}^2}\left(
(1 + \bar{H}^2 \tilde{u} \tilde{v})^2 - \frac{2 \kappa}{(\beta -2)(\beta-1)} (- \tilde{u})^{2 - \beta} + \cdots 
\right) d \Omega^2,
\end{align}
where $\tilde{u} \in (- \infty, 0)$ and $\tilde{v} \in (- \infty, \infty)$.

By identifying $u < 0$ as $\tilde{u}$ and $v$ as $\tilde{v}$, we can joint the expanding and contracting universe as
\begin{align}
 g_{\mu\nu}dx^{\mu} dx^{\nu} & =  - 2 d u d v +  \bar{H}^2 |u|^2 \Bigl(1  - \frac{2 \kappa }{(\beta - 3) ( \beta -2)} |u|^{2 - \beta} + \cdots \Bigr) d v^2 \notag\\
& \qquad  + \frac{1}{\bar{H}^2}\left(
(1 + \bar{H}^2 u v)^2 - \frac{2 \kappa}{(\beta -2)(\beta-1)} |u|^{2 - \beta} + \cdots 
\right) d \Omega^2.\label{completeFLRW} 
\end{align} 
The curvature component $A(u)$ is given by
\begin{align}
 A(u) = - \frac{\kappa}{|u|^{\beta}}.
\end{align}  

\subsection{Penrose limit of FLRW Universe}
\label{sec23}
The purpose of this paper is to discuss the property of the past boundary of inflationary universe by investigating a test string on the past completed inflationary universe \eqref{completedgeneralFLRW}. However, it is hard to perform the quantization of the string directly on the FLRW universe, because the wave equations is in general nonlinear equation of motion. A way to see the effect of past boundary is taking the Penrose limit~\cite{Penrose1976}, which is the limit to expand an infinitesimal neighborhood of a null geodesic outward. As a result, a general spacetime reduces to a plane wave spacetime.

Let us take the Penrose limit of \eqref{completedgeneralFLRW} with respect to a null geodesic $v = 0$.
By introducing new coordinates $(u, \hat{v}, x, y)$ as $ v = \epsilon^2 \hat{v}, \theta = \theta_{0} + \epsilon x, \phi = \phi_{0} + \frac{1}{\sin \theta_{0}} \epsilon y$, the metric \eqref{completedgeneralFLRW} can be expressed as
\begin{align}
 g_{\mu\nu}dx^{\mu}dx^{\nu} &= - 2 \epsilon^2 du d \hat{v} + a^2(|u|) \epsilon^4 d\hat{v}^2 \notag\\
& \qquad + a^2(|u|) (\text{sgn}(u) \eta(|u|) - \epsilon^2 \hat{v})^2 \epsilon^2 \left(d x^2 + \frac{\sin^2 (\theta_{0} + \epsilon x)}{\sin^2 \theta_{0}} d y^2 \right).
\end{align}
The Penrose limit is then defined by $g^{P}_{\mu\nu}dx^{\mu} dx^{\nu} := \lim_{\epsilon \to 0} \frac{1}{\epsilon^2} g_{\mu\nu} dx^{\mu} dx^{\nu} $. Thus we obtain,
\begin{align}
 g^{P}_{\mu\nu} dx^{\mu} dx^{\nu} = - 2 d u d \hat{v} + a(|u|)^2 \eta(|u|)^2 \left( d x^2 + d y^2 \right).
\end{align}
This is a plane wave spacetime in the Rosen coordinates. To discuss the string propagation in the next section, we will use other coordinates called the Brinkmann coordinates. The coordinate transformation to the Brinkmann coordinates is given by
\begin{align}
 x = \frac{1}{a \eta} X, \qquad y= \frac{1}{a \eta} Y, \qquad  \hat{v} = V - \frac{1}{2} \frac{(a \eta)'}{a \eta} (X^2 + Y^2),
\end{align}
and we obtain the expression
\begin{align}
 g^{P}_{\mu\nu} dx^{\mu} dx^{\nu} = - 2 d u d V + A(u) (X^2 + Y^2) du^2 + dX^2 + dY^2
\label{Brinkmann}
\end{align}
where $A(u)$ is defined by Eq.~\eqref{Au} in each coordinate patch.

Note that any of curvature invariants vanishes for the plane wave,
\begin{align}
 R^{P} = R^{P}_{\mu\nu} R^{P}{}^{\mu\nu} = R^{P}_{\mu\nu\rho\sigma} R^{P}{}^{\mu\nu\rho\sigma} = \cdots = 0.
\end{align}
Nonetheless, the curvature tensor has a non-vanishing component, 
\begin{align}
R^{P}_{\mu\nu} dx^{\mu} dx^{\nu} = - 2 A(u) du^2 .
\end{align}
As clear from this expression, the plane wave spacetime $g_{\mu\nu}^{P}$ has a parallelly propagated curvature singularity if the original universe~\eqref{completedgeneralFLRW} does. Thus we can discuss the effect of this singularity even after taking the Penrose limit.  It is worth to mention that for the example \eqref{A(u)=} the behavior of $A(u)$ is beyond the universality of Penrose limit proposed in \cite{Blau:2003dz,Blau:2004yi}, where the Penrose limit of a physically reasonable spacetime with a singularity is conjectured to have $A(u) \propto u^{-2}$.

\section{Excitation of string oscillation modes  on plane wave spacetime}\label{sec3}
In the previous section, we obtain the plane wave metric \eqref{Brinkmann} by taking the Penrose limit of past completed inflation universe \eqref{completedgeneralFLRW}. Excitation of oscillation modes of a string passing through a plane wave singularity has been studied in Refs.~\cite{Horowitz:1989bv,deVega:1990kk,Horowitz:1990sr,deVega:1990ke,David:2003vn}. Here we revisit this analysis with aiming an application to our inflationary universe. In particular, we will clarify the relation between the strength of the singularity and the excitation of string oscillations.

\subsection{Equations of motion of a string}
Let us consider the Polyakov action with four dimensional curved target space,
\begin{align}
S[h^{ab}, X^{\mu}] := -\frac{1}{4\pi\alpha'}\int d\tau d\sigma \sqrt{- h}h^{ab}g_{\mu\nu}(X)\partial_{a}X^\mu \partial_{b}X^\nu,
\end{align}
where $X^{\mu}$ $(\mu = 0, 1, 2, 3)$ are the embeddings of the world sheet on the spacetime, $h_{ab}$ $(a, b  = 0, 1)$ is the world sheet metric and $\alpha'$ is the Regge slope parameter. $g_{\mu\nu}(X)$ is the metric of the target space, which we identify as an inflation spacetime \eqref{completedgeneralFLRW} or the plane wave spacetime \eqref{Brinkmann} after taking the Penrose limit. The complete set of the equations of motion, in the conformal gauge $h_{ab}=e^{2\omega(\tau,\sigma)}\eta_{ab}$ with the world sheet Minkowski metric $\eta_{ab}$, are given by
\begin{align}
\left. \frac{- 4 \pi \alpha'}{\sqrt{- h}}  \frac{\delta S}{\delta X^{\mu}} \right|_{h = \mathrm{e}^{2\omega} \eta} &= \eta^{ab} \partial_{a} \partial_{b} X^\mu + \eta^{ab} \Gamma^\mu_{\nu\lambda}(X) \partial_a X^\nu \partial_b X^\lambda =0, \label{eomX}\\ 
\left. \frac{- 4\pi \alpha'}{\sqrt{- h}}\frac{\delta S}{\delta h^{ab}}\right|_{h = \mathrm{e}^{2\omega} \eta} &= 
g_{\mu\nu} \partial_{a} X^{\mu} \partial_{b} X^{\nu} - \frac{1}{2} \eta_{ab} \eta^{cd} g_{\mu\nu} \partial_{c} X^{\mu} \partial_{d} X^{\nu} =0. \label{eomh}
\end{align}
Since the trace part of the Eqs.~\eqref{eomh} vanishes automatically, only two components are independent.  

Let us solve the equations of motion by taking the Penrose limit. Plugging the plane wave metric \eqref{Brinkmann} into the equations of motion, and fixing the remaining gauge degree of freedom by the light-cone gauge $u = \alpha' p \tau$, we obtain following wave equations for $X^{i} = \{X, Y\}$,
\begin{align}
 &  \left(- \partial_{\tau}^2  + \partial_{\sigma}^2 + \alpha'{}^2 p^2 A(\alpha' p \tau) \right) X^{i}(\tau, \sigma) = 0,  \label{eomXi}
\end{align}
and for $V$,
\begin{align}
 (-\partial^2_{\tau} + \partial_{\sigma}^2) V(\tau,\sigma) = -\frac{\alpha'p}{2} \partial_{\tau}  \left(A(\alpha' p \tau)X^i X_i\right) -\alpha'p A(\alpha' p \tau) X^i \partial_{\tau} X_i   \label{eomV}
\end{align}
as well as two equations from Eq~\eqref{eomh},
\begin{align}
\alpha'p \partial_\tau V(\tau,\sigma)&=\frac{1}{2}\alpha'^2 p^2 A(\alpha'p\tau)X^i X_i+\frac{1}{2}(\partial_\sigma X^i)^2+\frac{1}{2}(\partial_\tau X^i)^2, \label{Vtau=}\\
\alpha'p \partial_\sigma V(\tau,\sigma)&=\partial_\tau X^i \partial_\sigma X^i. \label{Vcond}
\end{align}
Eqs.~\eqref{eomV}, \eqref{Vtau=} and \eqref{Vcond} can be solved by  
\begin{align}
 V(\tau, \sigma) = v_0 + \frac{1}{\alpha'p}\int^{\sigma} d\sigma \partial_{\tau} X^{i} \partial_{\sigma} X^{i} ,
\label{V=}
\end{align}
where $v_0$ is constant. Thus Eq. \eqref{V=}, as well as the light-cone gauge condition $u = \alpha' p \tau$, reduces the dynamical variables to $X^{i} = \{X, Y\}$.  The important point here is that the remaining equations of motion \eqref{eomXi} are linear in $X^{i}$ and we can quantize this system by similar way to the flat spacetime.
In particular, we can separately treat each Fourier modes of $X^{i}$,
\begin{align}
 X^{i}(\sigma, \tau) = X_{0}(\tau) + \sum_{n = 1}^{\infty} \left( X^{i}_{n}(\tau) \mathrm{e}^{i n \sigma} + X^{i *}_{n}(\tau) \mathrm{e}^{- i n \sigma} \right). 
\end{align}
By plugging this expression into \eqref{eomXi}, we obtain
\begin{align}
 - \frac{d^2 X^{i}_{n}}{d \tau^2} + {\cal V}(\tau) X^{i}_{n}  =  E_{n} X^{i}_{n} , \quad {\cal V}(\tau) := \alpha'{}^2 p^2 A(\alpha' p \tau), \quad  E_{n} := n^2 .
\label{Scheq}
\end{align}
This is nothing but the Schr\"{o}dinger equation with the potential ${\cal V}(\tau)$ and the energy $E_{n}$. 

\subsection{Excitation of oscillation modes}\label{excitation}
\begin{figure*}
\centering
 \includegraphics[width=0.5\hsize]{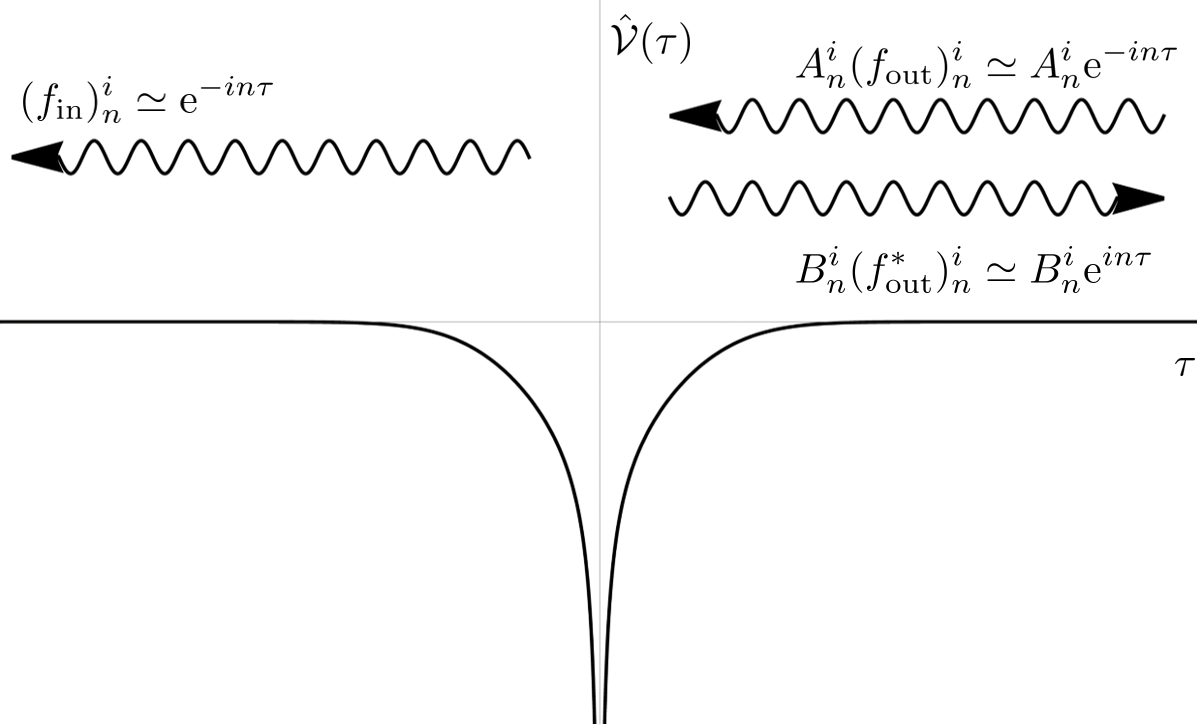}
 \caption{The replaced potential $\hat{{\cal V}}$ for an example $A(u) = - \kappa/|u|^{\beta}$  and  boundary conditions to obtain the Bogoliubov coefficients. }
\label{fig3}
\end{figure*}

Let us evaluate the excitation of oscillation modes of a string passing through $u = 0$. To pick up the effect of a singularity at $u = 0$, we replace $A(u)$ with $\hat{A}(u)$ defined by
\begin{align}
 \hat{A}(u) :=A(u)e^{- C u^2/(\alpha' p)^2},
\end{align}
where $C$ is a positive constant, and solve the Schr\"{o}dinger equation \eqref{Scheq} with the replaced potential $\hat{{\cal V}}(\tau):= \alpha'{}^2 p^2 \hat{A}(\alpha' p \tau)$ (See Fig. \ref{fig3}). 
Since the replaced potential dumps rapidly in each asymptotic region $\tau \rightarrow \pm \infty$, our Schr\"{o}dinger equation has positive and negative frequency solution $\mathrm{e}^{\pm i n \tau}$ there. We define $(f_{\text{in}})^{i}_{n}$ by the solution of Schr\"{o}dinger equation with the boundary condition $(f_{\text{in}})^{i}_{n} \overset{\tau \rightarrow - \infty}{\simeq} \mathrm{e}^{- i n \tau}$, while $(f_{\text{out}})^{i}_{n}$ by that with the boundary condition $(f_{\text{out}})^{i}_{n} \overset{\tau \rightarrow + \infty}{\simeq} \mathrm{e}^{- i n \tau}$.
By using these basis, general solution of the Schr\"{o}dinger equation can be written as following two ways,
\begin{align}
 X^i_n(\tau)&= 
i \frac{\sqrt{\alpha'}}{n}\biggl((\alpha_{\text{in}})^i_{n} (f_{\text{in}})^{i}_{n}(\tau) - (\tilde{\alpha}_{\text{in}}^{\dag})^i_{n} (f_{\text{in}}^{*})^{i}_{n}(\tau) \biggr)
\notag\\
&= i \frac{\sqrt{\alpha'}}{n}\biggl((\alpha_{\text{out}})^i_{n} (f_{\text{out}})^{i}_{n}(\tau) - (\tilde{\alpha}_{\text{out}}^{\dag})^i_{n} (f_{\text{out}}^{*})^{i}_{n}(\tau) \biggr).
\end{align}
Each mode operators $(\alpha_{\text{in}})^i_{n}$, $(\tilde{\alpha}_{\text{in}})^i_{n}$, $(\alpha_{\text{out}})^i_{n}$, $(\tilde{\alpha}_{\text{out}})^i_{n}$ satisfies the canonical commutation relation,
\begin{align}
 [(\alpha_{\text{in}})^{i}_{n} , (\alpha_{\text{in}}^{\dagger})^{j}_{m}] = n \delta_{n m} \delta^{ij},
\end{align}
and so on.

We consider the quantum vacuum $\ket{0_{\text{in}}}$ with no oscillation modes in the initial time:
\begin{align}
 (\alpha_{\text{in}})_{n}^{i} \ket{0_{\text{in}}} =  (\tilde{\alpha}_{\text{in}})_{n}^{i} \ket{0_{\text{in}}} = 0.
\end{align}
To discuss the excitation of oscillation modes, let us define Bogoliubov coefficients by
\begin{align}
 (f_{\text{in}})^{i}_{n}(\tau) &= A^{i}_{n} (f_{\text{out}})^{i}_{n}(\tau) + B^{i}_{n} (f_{\text{out}}^{*})^{i}_{n}(\tau), \notag\\
 (f_{\text{in}}^{*})^{i}_{n}(\tau) &= A^{i}_{n}{}^{*} (f_{\text{out}}^{*})^{i}_{n}(\tau) + B^{i}_{n}{}^{*} (f_{\text{out}})^{i}_{n}(\tau).
\end{align}
The operators $(\alpha_{\text{out}})^i_{n}$, $(\tilde{\alpha}_{\text{out}})^i_{n}$ are related to $(\alpha_{\text{in}})^i_{n}$, $(\tilde{\alpha}_{\text{in}})^i_{n}$ by the Bogoliubov transformation
\begin{align}
(\alpha_{\text{out}})^i_{n} &= A^i_n (\alpha_{\text{in}})^i_{n} - B^i_n{}^{*} (\tilde{\alpha}_{\text{in}}^{\dag})^i_{n},\\
(\tilde{\alpha}_{\text{out}})^i_{n} &=A^i_{n} (\tilde{\alpha}_{\text{in}})^i_{n} -  B^i_{n}{}^{*} (\alpha_{\text{in}}^{\dag})^i_{n}.
\end{align} 
The mass squared in the region $\tau \rightarrow \infty$ is defined by
\begin{align}
M^2_{\text{out}}&=\frac{1}{\alpha'} \sum^{\infty}_{n=1} \sum_{i = 1}^{2} \left(
(\alpha_{\text{out}}^{\dag})^{i}_{n}(\alpha_{\text{out}})^i_{n}+(\tilde{\alpha}_{\text{out}}^{\dag})^{i}_{n}(\tilde{\alpha}_{\text{out}})^i_{n}
\right)+m_0^2,
\end{align} 
where $m_{0}^2$ is the zero point energy. Therefore, the expectation value in our vacuum state $\ket{0_{\text{in}}}$ is expressed by the Bogoliubov coefficients as
\begin{align}
\langle M^2_{\text{out}}\rangle &=\frac{\bra{0_{\text{in}}} M^2_{\text{out}} \ket{0_{\text{in}}}}{\braket{0_{\text{in}}|0_{\text{in}}}}= m_0^2+\frac{2}{\alpha'}\sum^{\infty}_{n=1}\sum_{i = 1}^{2}n|B_n^i|^2  \label{M2}.
\end{align} 

Let us derive a useful formula to express the Bogoliubov coefficients $B_{n}^{i}$. By using the Wronskian
\begin{align}
 W[f,g] := \frac{d f}{d \tau} g - f \frac{d g}{d \tau},
\end{align} 
one can write the Bogoliubov coefficients $B_{n}^{i}$ as
\begin{align}
 B_{n}^{i} = \frac{W[ (f_{\text{out}})^{i}_{n}, (f_{\text{in}})^{i}_{n}]}{W[(f_{\text{out}})^{i}_{n}, (f^{*}_{\text{out}})^{i}_{n}]}.
\end{align}
Since the Wronskian is constant in time $\tau$, we can evaluate the value in the limit $\tau \rightarrow + \infty$, 
\begin{align}
 W[(f_{\text{out}})^{i}_{n}, (f^{*}_{\text{out}})^{i}_{n}] &= - 2 i n,\\
 W[(f_{\text{out}})^{i}_{n}, (f_{\text{in}})^{i}_{n}] &=
\lim_{\tau \rightarrow \infty} \left( - i n \mathrm{e}^{- i n \tau} (f_{\text{in}})^{i}_{n} - \mathrm{e}^{- i n \tau} \frac{d(f_{\text{in}})^{i}_{n}}{d \tau} \right) \notag\\
& = - \int^{\infty}_{- \infty} d\tau \mathrm{e}^{- i n \tau} \left( \frac{d^2 (f_{\text{in}})^{i}_{n}}{d \tau^2} + n^2 (f_{\text{in}})^{i}_{n} \right)
 \notag\\
&= - \alpha'{}^2 p^2 
\int^{\infty}_{- \infty} d \tau \mathrm{e}^{- i n \tau} \hat{A}(\alpha' p \tau) (f_{\text{in}})^{i}_{n}(\tau)
\end{align}
Thus, 
we obtain the expression of $B^{i}_{n}$ as follows, 
\begin{align}\label{B}
B^i_n=\frac{p^2\alpha'^2}{2in}
\int^{\infty}_{- \infty}d\tau\ \mathrm{e}^{-in\tau} (f_{\text{in}})^i_n(\tau) \hat{A}(\alpha'p\tau).
\end{align} 
Under the Born approximation $(f_{\text{in}})^i_n(\tau)\simeq \mathrm{e}^{-in\tau}$, $B_{n}^{i}$ can be evaluated as a Fourier transformation of the potential $\hat{A}$,  
\begin{align}
B^i_n \simeq \frac{p^2\alpha'^2}{2in}
\int^{\infty}_{- \infty}d\tau\ \mathrm{e}^{-2 i n \tau} \hat{A}(\alpha'p\tau). \label{BBorn}
\end{align} 

\subsection*{An example}
Again let us consider the example \eqref{a(u)=}, where the potential $A(u)$ is given by $A(u) = - \kappa/|u|^\beta$. Then we obtain
\begin{align}
B^i_n& = \frac{p^2 \alpha'^2}{2in} \int^{\infty}_{- \infty} d\tau\ \mathrm{e}^{-2 i n\tau} \frac{(- \kappa)}{|\alpha'p \tau|^{\beta}}\mathrm{e}^{- C \tau^2}
\notag\\
& \overset{n \rightarrow \infty}{\sim}
\begin{cases}
 {\cal B} n^{\beta-2} & (\beta < 1) \\
 \infty & (1 \leq \beta < 2) 
\end{cases}
,
\end{align}
where the coefficient ${\cal B}$ is independent from the cut off parameter $C$ and given by
\begin{align}
 {\cal B} = i\kappa 2^{\beta -1} (\alpha'p)^{2-\beta}\sin \left(\frac{\pi  \beta}{2} \right) \Gamma (1-\beta ).
\end{align}
For $\beta < 1$, $\braket{M^2_{\text{out}}}$ behaves as
\begin{align}
 \braket{M^2_{\text{out}}} \overset{n \rightarrow \infty}{\sim}
 \sum_n n^{2\beta-3},
\end{align} 
that is finite because the infinite sum of $n^{-s}$ is finite for $s > 1$.
Since $B_{n}^{i}$ itself are diverge for $1 \leq \beta < 2$, the behavior of the mass squared can be summarized as
\begin{align}
\langle M^2_{\text{out}}\rangle =
\begin{cases}
\text{finite} & \text{for $\beta<1$},\\
\text{divergent} & \text{for $1\leq\beta<2$} .
\end{cases}
\label{massdiv}
\end{align} 
Thus, the propagation of a string is well defined only when $\beta < 1$. In other words, The excitation of infinitely heavy modes can be avoidable only when the past boundary is a regular ($\beta < 0$) or weak singularity ($0 < \beta < 1$) in the sense of Krolak's classification. 

Before closing this section, we give some comments on the behavior for $1 \leq \beta < 2$ case. Since the above analysis is based on the Born approximation, one may think that our result, $B_{n}^{i} \rightarrow \infty$, is possibly an artifact of this approximation. To clarify this point, let us reveal an ill-behavior of the Schr\"{o}dinger equation for $1 \leq \beta$ case without using the approximation. As the discussion of the junction condition for wave function at a point where the potential has discontinuity, the $\tau$ integration of our Schr\"{o}dinger equation \eqref{Scheq} gives the following condition,
\begin{align}
 \left( \frac{d X_{n}^{i} (+ 0)}{d \tau} - \frac{d X_{n}^{i} (- 0)}{d \tau}\right)  = \lim_{\delta \rightarrow 0} \int_{-\delta}^{+\delta} d\tau\ (\mathcal{V}(\tau) - E_{n} ) X_{n}{}^{i} (\tau).
\end{align}
To obtain a well-defined solution, the right hand side must be finite.
For $1\leq\beta$ case, the right hand side diverges unless $X_{n}^{i} \overset{\tau \rightarrow 0}{\sim} 0$.
Since our boundary condition $X_{n}^{i} \overset{\tau \rightarrow - \infty}{\sim} \mathrm{e}^{ - i n \tau}$ has nothing to do with the condition $X_{n}^{i} \overset{\tau \rightarrow 0}{\sim} 0$ in general, we can conclude that this system is ill-defined for $1 \leq \beta$. As an example, let us focus on the case $\beta = 1$ case, where the Schr\"{o}dinger equation \eqref{Scheq} can be solved exactly in terms of hypergeometric function.  One can check that the solution which satisfied the condition $X_{n}^{i} \overset{\tau \rightarrow 0}{\sim} 0$ has the asymptotic property,
\begin{align}
 X_{n}^{i} (\tau) \overset{\tau \rightarrow - \infty}{\sim}  \frac{(- 2 i \tau)^{i \alpha' p \kappa /2n}}{\Gamma(1 + i \alpha' p \kappa/2n)} \mathrm{e}^{- i n \tau} - \frac{(2 i \tau)^{- i \alpha' p \kappa/2n}}{\Gamma(1 - i \alpha' p \kappa/2n)} \mathrm{e}^{i n \tau},
\end{align}
which is a mixing of positive and negative frequency modes.

\section{String Excitation by initial singularity of inflation}\label{sec4}
Let us apply the above analysis to various models of single field inflation and see the consistency of the effective field theory description. We consider the Einstein--Hilbert action with an inflaton $\phi$ with a potential $V(\phi)$,
\begin{align}
S=\int d^4 x\sqrt{-g}\left(\frac{M_{pl}^2}{2} R - \frac{1}{2}g^{\mu\nu}\partial_\mu\phi \partial_\nu\phi
-V(\phi)
\right).
\end{align} 
Assuming the flat FLRW ansatz for the metric \eqref{FLRW} and the homogeneity of inflaton $\phi = \phi(t)$, this action yields following equations of motion,
\begin{align}
& \ddot{\phi} +3H \dot{\phi}+ V'(\phi)=0, \qquad   H^2=\frac{1}{3M_{pl}^2}\left(\frac{1}{2} \dot{\phi}^2+V(\phi)\right).
\end{align}  
Under the slow roll condition $|\ddot{\phi}|\ll 3 H |\dot{\phi}|$ and $\dot{\phi}^2/2 \ll V(\phi)$, the equations of motion reduce to
\begin{align}
 H \simeq \sqrt{\frac{V(\phi)}{3 M_{pl}^2}}, \qquad \dot{\phi} \simeq - \frac{V'(\phi)}{3 H}.
\end{align}
By using equations of motion, slow roll condition can be rewrite as $\epsilon \ll 1$ and $|\eta| \ll 1$ with 
\begin{align}
 \epsilon := \frac{M_{pl}^2}{2} \left(\frac{V'}{V}\right)^2,\qquad  \eta:= M_{pl}^2 \frac{V''}{V}.
\end{align}
To see the strength of the singularity, we need to evaluate $A = \dot{H}/a^2$ as a function of $u$. Recall that we are interested in inflationary solutions with the property $ a \simeq \mathrm{e}^{\bar{H} t}$ in the limit $t \rightarrow -\infty$. Then we rewrite the potential $V(\phi)$ as
\begin{align}
 V(\phi) = 3 M_{pl}^2 \bar{H}^2 \left( 1 + \delta V(\phi) \right),
\end{align}
and we focus on specific solutions
\begin{align}
 \lim_{\phi \rightarrow \phi_{- \infty}} \delta V(\phi) = 0, \qquad \lim_{t \rightarrow -\infty} \phi = \phi_{- \infty}.
\end{align}

As mentioned before, the leading term of $a$ can be expressed as
\begin{align}
 a = \bar{H} u + {\cal O}(u^2). \label{scaleleading}
\end{align}
On the other hand, by the equations of motion under the slow roll condition, $a$ can be expressed as a function of $\phi$ as $a = a_{e}\mathrm{e}^{- {\cal N}(\phi)}$, where $a_{e}$ is a constant and ${\cal N}(\phi)$ is the e-folding number defined by
\begin{align}
 {\cal N}(\phi) := \frac{1}{M_{pl}^2} \int^{\phi}_{\phi_{e}} d \phi \frac{V(\phi)}{V'(\phi)} \simeq \frac{1}{M_{pl}^2} \int^{\phi}_{\phi_{e}} d \phi \frac{1}{\delta V'(\phi)}. \label{e-fold}
\end{align}
Here we assume $1 + \delta V \sim 1$ in the final expression.
By comparing two expressions, we can obtain a relation between $u$ and $\phi$,
\begin{align}
\log \frac{\bar{H} u}{a_{e}} = - \frac{1}{M_{pl}^2}\int^{\phi}_{\phi_{e}} \frac{d \phi}{\delta V'(\phi)} ,\label{phi=phi(u)}
\end{align}
which can be used to express $\phi$ as a function of $u$. 

For the equations of motion under the slow roll condition, we can write $dH/dt$ as a function of $\phi$, 
\begin{align}
 \frac{dH(t)}{dt} &= - \frac{1}{6} \frac{V'(\phi)^2}{V(\phi)} \simeq - \frac{M_{pl}^2 \bar{H}^2}{2} \delta V'(\phi(u))^2 .
\end{align}
Thus the component of the curvature tensor, $A(u)$, can be expressed as
\begin{align}
A(u) = \frac{1}{a^2} \frac{d H}{dt} \simeq - \frac{M_{pl}^2}{2u^2}\times\delta V'(\phi(u))^2. \label{infA}
\end{align}
To summarize, for a given inflaton potential $V(\phi)$, we can get the expression $\phi = \phi (u)$ from Eq.\eqref{phi=phi(u)}, and then we can obtain $A(u)$ from \eqref{infA}.

\subsection{Starobinsky type model}
Let us consider a Starobinsky-type inflation model with a potential
\begin{align}
 V(\phi) &= 3 M_{pl}^2 \bar{H}^2 \left( 1 - \mathrm{e}^{- \frac{\phi}{\mu}} + \cdots \right),
\end{align}
in the large field region $\phi \gg \mu$. This model includes the $\alpha$ attractor model \cite{Ferrara:2013rsa,Kallosh:2013yoa,Carrasco:2015pla} and pole inflation model \cite{Galante:2014ifa,Broy:2015qna,Terada:2016nqg,Saikawa:2017wkg} in with a pole of order 2. 
We are interested in the solution that the slow roll conditions are satisfied $t \rightarrow - \infty$, otherwise the kinetic energy of inflaton dominates over the vacuum energy in the past \cite{Tsamis:2003px,Kinney:2005vj,Cai:2016ngx}, which leads to a scalar curvature singularity.

We can evaluate $\delta V'$ as
\begin{align}
 \delta V'(\phi) = \frac{1}{\mu} \mathrm{e}^{- \frac{\phi}{\mu}}.
\end{align}  
Then Eq.~\eqref{phi=phi(u)} can be evaluated as
\begin{align}
 \log \frac{\bar{H} u}{a_{e}} = - \frac{\mu}{M_{pl}^2} \int^{\phi}_{\phi_{e}} d\phi\, \mathrm{e}^{\frac{\phi}{\mu}}
= - \frac{\mu^2}{M_{pl}^2}\left( \mathrm{e}^{\frac{\phi}{\mu}} - \mathrm{e}^{\frac{\phi_{e}}{\mu}}\right).
\end{align}
Thus we can write $\phi$ as a function of $u$ by
\begin{align}
 \mathrm{e}^{\frac{\phi}{\mu}} = \mathrm{e}^{ \frac{\phi_{e}}{\mu}} - \frac{M_{pl}^2}{\mu^2} \log \frac{\bar{H} u}{a_{e}},
\end{align}
and $A(u)$ can be represented as
\begin{align}
 A(u) &= - \frac{M_{pl}^2}{2 u^2} \times \frac{1}{\mu^2} \left( \mathrm{e}^{ \frac{\phi_{e}}{\mu}} - \frac{M_{pl}^2}{\mu^2} \log \frac{\bar{H} u}{a_{e}} \right)^{-2} \notag\\
&= - \frac{\kappa}{u^2 \left(c_{1} - c_2 \log \frac{\bar{H} u}{a_{e}}\right)^2 },
\end{align}
with constants $\kappa, c_1$ and $c_2$.
We observe that the divergence of $A(u)$ in the limit $u \rightarrow 0$ is stronger than $1/u$, while it is weaker than $1/u^2$.
Therefore, we conclude that this is a strong singularity in the sense of Krolak and the infinitely heavy oscillation modes are excited.

\subsection{General hill top inflation}\label{hilltop}
Let us consider a general hill top potential,
\begin{align}
 V(\phi) = 3 M_{pl}^2 \bar{H}^2 \left( 1 - \left( \frac{\phi}{\mu}\right)^{n} + {\cal O}\left(\left(\frac{\phi}{\mu} \right)^{n+1} \right) \right), \qquad n > 2. 
\end{align}
Then the slow roll conditions are satisfied in the limit $\phi \rightarrow 0$.
We focus on the specific solution that the inflaton had been rolling down the potential in an infinite time: $\lim_{t \rightarrow - \infty} \phi = 0$.

The derivative of $\delta V$ can be evaluated as $\delta V'(\phi) = - n \phi^{n-1}/\mu^{n}$ and hence Eq.~\eqref{phi=phi(u)} reduces to
\begin{align}
  \log \frac{\bar{H} u}{a_{e}} = \frac{1}{M_{pl}^2} \frac{\mu^{n}}{n}\int^{\phi}_{\phi_{e}} \frac{d \phi}{\phi^{n-1}} =  - \frac{1}{n(n - 2)} \frac{\mu^2}{M_{pl}^2} \left( \left(\frac{\mu}{\phi} \right)^{n - 2} - \left( \frac{\mu}{\phi_{e}} \right)^{n - 2}\right)
\end{align}
Thus $\phi$ can be represented as 
\begin{align}
  \frac{\phi}{\mu} = \left( \left(\frac{\mu}{\phi_{e}}\right)^{n-2}  -  n(n-2) \frac{M_{pl}^2}{\mu^2} \log \frac{\bar{H} u}{a_{e}} \right)^{- \frac{1}{n-2}}.
\end{align}
Therefore, $A(u)$ can be evaluated as
\begin{align}
 A(u) &= - \frac{M_{pl}^2}{2 u^2} \times \frac{n^2}{\mu^2} \left( \left(\frac{\mu}{\phi_{e}}\right)^{n-2}  -  n(n-2) \frac{M_{pl}^2}{\mu^2} \log \frac{\bar{H} u}{a_{e}} \right)^{- \frac{2n - 2}{n-2}} \notag\\
 &= - \frac{\kappa}{u ^2 \left(c_1  - c_2 \log \frac{\bar{H} u}{a_{e}}\right)^{\frac{2n-2}{n-2}}}.
\end{align}
Again, the divergence of $A(u)$ is stronger than $1/u$. Hence, we conclude that this is a strong singularity in the sense of Krolak and the infinitely heavy oscillation modes are excited. 

\subsection{Quadratic hill top Inflation}
\label{sec:4.3}
Let us consider a slow roll inflation caused by an inflaton with a quadratic hill top potential,
\begin{align}
 V(\phi) = 3 M_{pl}^2 \bar{H}^2 \left( 1 - \frac{|\bar{\eta}|}{2} \frac{\phi^2}{M_{pl}^2} + {\cal O}\left(\left(\frac{\phi}{M_{pl}} \right)^3 \right) \right),
\end{align}
where $\bar{\eta}$ is a negative parameter which represents the slow roll parameter $\eta$ in the limit $\phi \rightarrow 0$. Actually the slow roll parameters can be evaluated as 
\begin{align}
 \epsilon = \frac{|\bar{\eta}|^2}{2} \frac{\phi^2}{M_{pl}^2} + {\cal O}\left( \left(\frac{\phi}{M_{pl}}\right)^3 \right),\qquad  \eta = - |\bar{\eta}| + {\cal O}\left( \frac{\phi}{M_{pl}} \right).
\end{align}
Let us focus on the case $|\bar{\eta}| \ll 1$ first. In this case, the slow roll conditions are satisfied and we can use Eq.~\eqref{phi=phi(u)} and Eq.~\eqref{infA}. Since $\delta V'  \sim - |\bar{\eta}| \phi/ M_{pl}^2$, Eq.\eqref{phi=phi(u)} reduces to
\begin{align}
  \log \frac{\bar{H} u}{a_{e}} = - \frac{1}{M_{pl}^2}  \frac{(- M_{pl}^2)}{|\bar{\eta}|} \int^{\phi}_{\phi_{e}} \frac{d \phi}{\phi} = \frac{1}{|\bar{\eta}|} \log \frac{\phi}{\phi_{e}},\qquad  \Leftrightarrow \qquad  \phi(u) = \phi_{e} \left(\frac{\bar{H}}{a_{e}} u \right)^{|\bar{\eta}|}.
\end{align}
Then from Eq.~\eqref{infA}, we can write $A$ as a function of $u$,
\begin{align}
 A(u) & \sim - \frac{M_{pl}^2}{2 u^2} \times 
\frac{|\bar{\eta}|^2}{M_{pl}^4} \phi_{e}^2 \left(\frac{\bar{H}}{a_{e}} u \right)^{2 |\bar{\eta}|} 
 = - \frac{\kappa}{u^{\beta}},
\end{align}
with
\begin{align}
 \beta := 2 (1 - |\bar{\eta}|), \qquad \kappa :=
 \frac{|\bar{\eta}|^2 \phi_{e}^2}{2 M_{pl}^2} \left(\frac{\bar{H}}{a_{e}}\right)^{2 |\bar{\eta}|}
.
\end{align}
Thus the curvature component $A(u)$ of the quadratic hill top inflation coincides with that of the example studied in the previous sections. For the parameter $|\bar{\eta}| \ll 1$, we can see that $1 < \beta < 2$ and $u = 0$ is a strong singularity in the sense of Krolak. Thus the propagation of a quantum string is ill-defined for the small slow roll parameter $|\bar{\eta}| \ll 1$.

We can perform similar analysis for a large $|\bar{\eta}|$, though this case is disfavored from observations.  Let us relax the first slow roll condition $|\ddot{\phi}| \ll 3 H |\dot{\phi}|$, but assume the second one $\dot{\phi}^2/2 \ll V$.  Considering again the specific solution which satisfies $\lim_{t \rightarrow - \infty} \phi = 0$, the equations of motion in the limit $t \rightarrow - \infty$ reduce to 
\begin{align}
 \ddot{\phi} + 3 \bar{H} \dot{\phi} - 3 |\bar{\eta}| \bar{H}^2 \phi \simeq 0, \qquad H \simeq \bar{H}.
\end{align}
Then, we obtain $a \propto \mathrm{e}^{\bar{H} t}$  and $\phi \propto \mathrm{e}^{\gamma \bar{H} t}$, where $\gamma$ is given by
\begin{align}
\gamma := \frac{-3 + \sqrt{9 + 12 |\bar{\eta}|}}{2},
\end{align}
and hence $A$ can be evaluated as
\begin{align}
 A = \frac{\dot{H}}{a^2} = - \frac{1}{2 M_{pl}^2} \frac{\dot{\phi}^2}{a^2} \propto \mathrm{e}^{ - 2 (1 - \gamma) \bar{H} t} \propto \frac{1}{u^{2 ( 1 - \gamma)}}.
\end{align} 
Thus, $\beta = 2 (1 - \gamma)$ and we can conclude that
\begin{alignat*}{4}
 0 &< |\bar{\eta}| \leq \frac{7}{12} & \quad & \Leftrightarrow & \quad & 1\leq \beta < 2 &\qquad & (\text{Krolak's strong singularity}),\\
 \frac{7}{12} &< |\bar{\eta}| \leq \frac{4}{3} &&\Leftrightarrow&& 0 < \beta < 1&& (\text{Krolak's weak singularity}),\\
 \frac{4}{3} &< |\bar{\eta}| &&\Leftrightarrow&&  \beta < 0&& (\text{regular}).
\end{alignat*}
Thus, the excitation of infinitely heavy strings are avoidable when $|\bar{\eta}| > 7/12$ and effective description of the inflation models with this parameter looks consistent. This is a sharp contrast to the slow roll case $|\bar{\eta}| \ll 1$.

\section{Summary and Discussion}
In the present paper, we showed that infinitely heavy oscillation modes are excited when a string passes through the initial non-scalar curvature singularity of the slow roll solutions, in Starobinsky-type inflation and the hill top models. This result can be understood that stringy corrections are inevitable in the very early stage of slow roll inflation models, even though the value of the Hubble parameter remains smaller than the string scale. 

In the end of Sec. \ref{sec:4.3}, we find that the excitation of infinitely heavy modes is avoidable if one assume the violation of slow roll condition, though this case is disfavored from the observation. This fact provides an interesting insight to the validity of low energy effective theory based on string theory. Let us regard the inflaton as an axion and consider a hill top inflation caused by the axion cosine potential given by
\begin{align}
 V(\phi) = \frac{3 M_{pl}^2 \bar{H}^2}{2} \left( 1 + \cos \frac{\phi}{f}\right) = 3 M_{pl}^2 \bar{H}^2 \left( 1 - \frac{1}{2} \frac{M_{pl}^2}{2 f^2} \frac{\phi^2}{M_{pl}^2} + \cdots \right),
\end{align}
where the slow roll parameter $|\bar{\eta}|$ can read $|\bar{\eta}| = M_{pl}^2/(2 f^2)$.
Then the condition for the effective description of very early stage of inflation to be well-defined can be written as
\begin{align}
 f < \sqrt{\frac{6}{7}} M_{pl}.
\end{align}
This relation looks similar with the axionic weak gravity conjecture $f < {\cal O}(M_{pl})$ \cite{Brown:2015iha,Montero:2015ofa,Rudelius:2015xta}, see also Ref.~\cite{vanBeest:2021lhn}. Thus it is interesting to clarify the relation of the violation of effective description of inflation models due to the spacetime singularity to that based on the general property of quantum gravity. 

We mainly focused on the non-scalar curvature singularity here. 
It is also known that yet another type of singularity occurs inflation spacetime when the spatial topology is nontrivial \cite{Ishibashi:1996ps, Numasawa:2019juw}. This can be understood as a de Sitter space analogue of well known Misner singularity \cite{1967rta1.book..160M}. For this singularity, an infinite number of winding modes of string, not oscillation modes, will be excited \cite{Berkooz:2004re}. Thus it is interesting to study the excitation of winding modes in inflationary case. Because of the difficulty to quantize strings directly in curved spacetime, it will be useful to use a method like the Penrose limit or develop an alternative method to treat winding modes in expanding universe such as the double field theory \cite{Siegel:1993xq,Hull:2009mi,Hohm:2010jy,Hohm:2010pp,Zwiebach:2011rg} and its massive extension \cite{Hohm:2016lim,Noumi:2020php}. 

In the present work, we evaluated the excitation of string at a future time far enough from the singularity. It may be interesting to evaluate string excitation near the singularity depending on time. Then we can evaluate the effect of back reaction to the spacetime and discuss the resolution of the initial singularity by string excitation.  We leave above interesting directions for future work.

\acknowledgments
We would like to thank Toshifumi Noumi and Kimihiro Nomura for helpful discussion. D. Y. is supported by JSPS KAKENHI Grant Numbers JP19J00294 and JP20K14469.

\bibliographystyle{JHEP}
\bibliography{ref}

\providecommand{\href}[2]{#2}\begingroup\raggedright\begin{thebibliography}{10}

\bibitem{Akrami:2018odb}
{\scshape Planck} collaboration, \emph{{Planck 2018 results. X. Constraints on
  inflation}}, \href{https://doi.org/10.1051/0004-6361/201833887}{\emph{Astron.
  Astrophys.} {\bfseries 641} (2020) A10}
  [\href{https://arxiv.org/abs/1807.06211}{{\ttfamily 1807.06211}}].

\bibitem{Borde:2001nh}
A.~Borde, A.H.~Guth and A.~Vilenkin, \emph{{Inflationary space-times are
  incompletein past directions}},
  \href{https://doi.org/10.1103/PhysRevLett.90.151301}{\emph{Phys. Rev. Lett.}
  {\bfseries 90} (2003) 151301}
  [\href{https://arxiv.org/abs/gr-qc/0110012}{{\ttfamily gr-qc/0110012}}].

\bibitem{Yoshida:2018ndv}
D.~Yoshida and J.~Quintin, \emph{{Maximal extensions and singularities in
  inflationary spacetimes}},
  \href{https://doi.org/10.1088/1361-6382/aacf4b}{\emph{Class. Quant. Grav.}
  {\bfseries 35} (2018) 155019}
  [\href{https://arxiv.org/abs/1803.07085}{{\ttfamily 1803.07085}}].

\bibitem{Hawking:1973uf}
S.W.~Hawking and G.F.R.~Ellis, \emph{{The Large Scale Structure of
  Space-Time}}, Cambridge Monographs on Mathematical Physics, Cambridge
  University Press (2, 2011),
  \href{https://doi.org/10.1017/CBO9780511524646}{10.1017/CBO9780511524646}.

\bibitem{Ellis:1977pj}
G.F.R.~Ellis and B.G.~Schmidt, \emph{{Singular space-times}},
  \href{https://doi.org/10.1007/BF00759240}{\emph{Gen. Rel. Grav.} {\bfseries
  8} (1977) 915}.

\bibitem{FernandezJambrina:2007sx}
L.~Fernandez-Jambrina, \emph{{Hidden past of dark energy cosmological models}},
  \href{https://doi.org/10.1016/j.physletb.2007.08.091}{\emph{Phys. Lett. B}
  {\bfseries 656} (2007) 9} [\href{https://arxiv.org/abs/0704.3936}{{\ttfamily
  0704.3936}}].

\bibitem{Fernandez-Jambrina:2016clh}
L.~Fern\'andez-Jambrina, \emph{{Initial directional singularity in inflationary
  models}}, \href{https://doi.org/10.1103/PhysRevD.94.024049}{\emph{Phys. Rev.
  D} {\bfseries 94} (2016) 024049}
  [\href{https://arxiv.org/abs/1606.07600}{{\ttfamily 1606.07600}}].

\bibitem{Nomura:2021lzz}
K.~Nomura and D.~Yoshida, \emph{{Past extendibility and initial singularity in
  Friedmann-Lema\^\i{}tre-Robertson-Walker and Bianchi I spacetimes}},
  \href{https://arxiv.org/abs/2105.05642}{{\ttfamily 2105.05642}}.

\bibitem{Combes:1993rw}
F.~Combes, H.J.~de~Vega, A.V.~Mikhailov and N.G.~Sanchez, \emph{{Multistring
  solutions by soliton methods in de Sitter space-time}},
  \href{https://doi.org/10.1103/PhysRevD.50.2754}{\emph{Phys. Rev. D}
  {\bfseries 50} (1994) 2754}
  [\href{https://arxiv.org/abs/hep-th/9310073}{{\ttfamily hep-th/9310073}}].

\bibitem{deVega:1993rm}
H.J.~de~Vega, A.L.~Larsen and N.G.~Sanchez, \emph{{Infinitely many strings in
  de Sitter space-time: Expanding and oscillating elliptic function
  solutions}}, \href{https://doi.org/10.1016/0550-3213(94)90643-2}{\emph{Nucl.
  Phys. B} {\bfseries 427} (1994) 643}
  [\href{https://arxiv.org/abs/hep-th/9312115}{{\ttfamily hep-th/9312115}}].

\bibitem{Noumi:2019ohm}
T.~Noumi, T.~Takeuchi and S.~Zhou, \emph{{String Regge trajectory on de Sitter
  space and implications to inflation}},
  \href{https://doi.org/10.1103/PhysRevD.102.126012}{\emph{Phys. Rev. D}
  {\bfseries 102} (2020) 126012}
  [\href{https://arxiv.org/abs/1907.02535}{{\ttfamily 1907.02535}}].

\bibitem{Lust:2019lmq}
D.~L\"ust and E.~Palti, \emph{{A Note on String Excitations and the Higuchi
  Bound}}, \href{https://doi.org/10.1016/j.physletb.2019.135067}{\emph{Phys.
  Lett. B} {\bfseries 799} (2019) 135067}
  [\href{https://arxiv.org/abs/1907.04161}{{\ttfamily 1907.04161}}].

\bibitem{Kato:2021rdz}
M.~Kato, K.~Nishii, T.~Noumi, T.~Takeuchi and S.~Zhou, \emph{{Spiky strings in
  de Sitter space}},  \href{https://arxiv.org/abs/2102.09746}{{\ttfamily
  2102.09746}}.

\bibitem{Parmentier:2021nwz}
K.~Parmentier, \emph{{Mass and spin for classical strings in $dS_3$}},
  \href{https://arxiv.org/abs/2102.10805}{{\ttfamily 2102.10805}}.

\bibitem{Sanchez:1989cw}
N.G.~Sanchez and G.~Veneziano, \emph{{Jeans Like Instabilities for Strings in
  Cosmological Backgrounds}},
  \href{https://doi.org/10.1016/0550-3213(90)90230-B}{\emph{Nucl. Phys. B}
  {\bfseries 333} (1990) 253}.

\bibitem{Gasperini:1990xg}
M.~Gasperini, N.G.~Sanchez and G.~Veneziano, \emph{{Highly unstable fundamental
  strings in inflationary cosmologies}},
  \href{https://doi.org/10.1142/S0217751X91001878}{\emph{Int. J. Mod. Phys. A}
  {\bfseries 6} (1991) 3853}.

\bibitem{Larsen:1995vr}
A.L.~Larsen and N.G.~Sanchez, \emph{{Strings in standard expanding FRW
  universes}}, \href{https://doi.org/10.1103/PhysRevD.54.2483}{\emph{Phys. Rev.
  D} {\bfseries 54} (1996) 2483}
  [\href{https://arxiv.org/abs/hep-th/9511069}{{\ttfamily hep-th/9511069}}].

\bibitem{deVega:1995bq}
H.J.~de~Vega and N.G.~Sanchez, \emph{{Lectures on string theory in curved
  space-times}}, {\emph{NATO Sci. Ser. C} {\bfseries 476} (1996) 11}
  [\href{https://arxiv.org/abs/hep-th/9512074}{{\ttfamily hep-th/9512074}}].

\bibitem{Tolley:2005us}
A.J.~Tolley, \emph{{String propagation through a big crunch/big bang
  transition}}, \href{https://doi.org/10.1103/PhysRevD.73.123522}{\emph{Phys.
  Rev. D} {\bfseries 73} (2006) 123522}
  [\href{https://arxiv.org/abs/hep-th/0505158}{{\ttfamily hep-th/0505158}}].

\bibitem{Penrose1976}
R.~Penrose, \emph{Any space-time has a plane wave as a limit},  in
  \emph{Differential Geometry and Relativity: A Volume in Honour of Andr{\'e}
  Lichnerowicz on His 60th Birthday}, M.~Cahen and M.~Flato, eds., (Dordrecht),
  pp.~271--275, Springer Netherlands (1976),
  \href{https://doi.org/10.1007/978-94-010-1508-0_23}{DOI}.

\bibitem{Blau:2002dy}
M.~Blau, J.M.~Figueroa-O'Farrill, C.~Hull and G.~Papadopoulos, \emph{{Penrose
  limits and maximal supersymmetry}},
  \href{https://doi.org/10.1088/0264-9381/19/10/101}{\emph{Class. Quant. Grav.}
  {\bfseries 19} (2002) L87}
  [\href{https://arxiv.org/abs/hep-th/0201081}{{\ttfamily hep-th/0201081}}].

\bibitem{Geroch:1969ca}
R.P.~Geroch, \emph{{Limits of spacetimes}},
  \href{https://doi.org/10.1007/BF01645486}{\emph{Commun. Math. Phys.}
  {\bfseries 13} (1969) 180}.

\bibitem{Blau:2002mw}
M.~Blau, J.M.~Figueroa-O'Farrill and G.~Papadopoulos, \emph{{Penrose limits,
  supergravity and brane dynamics}},
  \href{https://doi.org/10.1088/0264-9381/19/18/310}{\emph{Class. Quant. Grav.}
  {\bfseries 19} (2002) 4753}
  [\href{https://arxiv.org/abs/hep-th/0202111}{{\ttfamily hep-th/0202111}}].

\bibitem{Blau:2003dz}
M.~Blau, M.~Borunda, M.~O'Loughlin and G.~Papadopoulos, \emph{{Penrose limits
  and space-time singularities}},
  \href{https://doi.org/10.1088/0264-9381/21/7/L02}{\emph{Class. Quant. Grav.}
  {\bfseries 21} (2004) L43}
  [\href{https://arxiv.org/abs/hep-th/0312029}{{\ttfamily hep-th/0312029}}].

\bibitem{Blau:2004yi}
M.~Blau, M.~Borunda, M.~O'Loughlin and G.~Papadopoulos, \emph{{The Universality
  of Penrose limits near space-time singularities}},
  \href{https://doi.org/10.1088/1126-6708/2004/07/068}{\emph{JHEP} {\bfseries
  07} (2004) 068} [\href{https://arxiv.org/abs/hep-th/0403252}{{\ttfamily
  hep-th/0403252}}].

\bibitem{Horowitz:1989bv}
G.T.~Horowitz and A.R.~Steif, \emph{{Space-Time Singularities in String
  Theory}}, \href{https://doi.org/10.1103/PhysRevLett.64.260}{\emph{Phys. Rev.
  Lett.} {\bfseries 64} (1990) 260}.

\bibitem{deVega:1990kk}
H.J.~de~Vega and N.G.~Sanchez, \emph{{QUANTUM STRING PROPAGATION THROUGH
  GRAVITATIONAL SHOCK WAVES}},
  \href{https://doi.org/10.1016/0370-2693(90)90058-E}{\emph{Phys. Lett. B}
  {\bfseries 244} (1990) 215}.

\bibitem{Horowitz:1990sr}
G.T.~Horowitz and A.R.~Steif, \emph{{Strings in Strong Gravitational Fields}},
  \href{https://doi.org/10.1103/PhysRevD.42.1950}{\emph{Phys. Rev. D}
  {\bfseries 42} (1990) 1950}.

\bibitem{deVega:1990ke}
H.J.~de~Vega and N.G.~Sanchez, \emph{{Strings falling into space-time
  singularities}}, \href{https://doi.org/10.1103/PhysRevD.45.2783}{\emph{Phys.
  Rev. D} {\bfseries 45} (1992) 2783}.

\bibitem{David:2003vn}
J.R.~David, \emph{{Plane waves with weak singularities}},
  \href{https://doi.org/10.1088/1126-6708/2003/11/064}{\emph{JHEP} {\bfseries
  11} (2003) 064} [\href{https://arxiv.org/abs/hep-th/0303013}{{\ttfamily
  hep-th/0303013}}].

\bibitem{Craps:2008bv}
B.~Craps, F.~De~Roo and O.~Evnin, \emph{{Can free strings propagate across
  plane wave singularities?}},
  \href{https://doi.org/10.1088/1126-6708/2009/03/105}{\emph{JHEP} {\bfseries
  03} (2009) 105} [\href{https://arxiv.org/abs/0812.2900}{{\ttfamily
  0812.2900}}].

\bibitem{Aichelburg:1970dh}
P.C.~Aichelburg and R.U.~Sexl, \emph{{On the Gravitational field of a massless
  particle}}, \href{https://doi.org/10.1007/BF00758149}{\emph{Gen. Rel. Grav.}
  {\bfseries 2} (1971) 303}.

\bibitem{Veneziano:1987df}
G.~Veneziano, \emph{{Mutual Focusing of Graviton Beams}},
  \href{https://doi.org/10.1142/S0217732387001142}{\emph{Mod. Phys. Lett. A}
  {\bfseries 2} (1987) 899}.

\bibitem{Amati:1988ww}
D.~Amati and C.~Klimcik, \emph{{Strings in a Shock Wave Background and
  Generation of Curved Geometry from Flat Space String Theory}},
  \href{https://doi.org/10.1016/0370-2693(88)90355-3}{\emph{Phys. Lett. B}
  {\bfseries 210} (1988) 92}.

\bibitem{deVega:1988ts}
H.J.~de~Vega and N.G.~Sanchez, \emph{{Quantum String Scattering in the
  Aichelburg-sexl Geometry}},
  \href{https://doi.org/10.1016/0550-3213(89)90540-3}{\emph{Nucl. Phys. B}
  {\bfseries 317} (1989) 706}.

\bibitem{deVega:1990nr}
H.J.~de~Vega and N.G.~Sanchez, \emph{{Space-time singularities in string theory
  and string propagation through gravitational shock waves}},
  \href{https://doi.org/10.1103/PhysRevLett.65.1517}{\emph{Phys. Rev. Lett.}
  {\bfseries 65} (1990) 1517}.

\bibitem{Sanchez:2003ek}
N.G.~Sanchez, \emph{{Classical and quantum strings in plane waves, shock waves
  and space-time singularities: Synthesis and new results}},
  \href{https://doi.org/10.1142/S0217751X03015787}{\emph{Int. J. Mod. Phys. A}
  {\bfseries 18} (2003) 4797}
  [\href{https://arxiv.org/abs/hep-th/0302214}{{\ttfamily hep-th/0302214}}].

\bibitem{Tipler:1977zza}
F.J.~Tipler, \emph{{Singularities in conformally flat spacetimes}},
  \href{https://doi.org/10.1016/0375-9601(77)90508-4}{\emph{Phys. Lett. A}
  {\bfseries 64} (1977) 8}.

\bibitem{krolak1983proof}
A.~Kr{\'o}lak, \emph{A proof of the cosmic censorship hypothesis},
  {\emph{General Relativity and Gravitation} {\bfseries 15} (1983) 99}.

\bibitem{CLARKE1985127}
C.~Clarke and A.~Kr{\'o}lak, \emph{Conditions for the occurence of strong
  curvature singularities},
  \href{https://doi.org/https://doi.org/10.1016/0393-0440(85)90012-9}{\emph{Journal
  of Geometry and Physics} {\bfseries 2} (1985) 127}.

\bibitem{Krolak_1986}
A.~Krolak, \emph{Towards the proof of the cosmic censorship hypothesis},
  \href{https://doi.org/10.1088/0264-9381/3/3/004}{\emph{Classical and Quantum
  Gravity} {\bfseries 3} (1986) 267}.

\bibitem{Ferrara:2013rsa}
S.~Ferrara, R.~Kallosh, A.~Linde and M.~Porrati, \emph{{Minimal Supergravity
  Models of Inflation}},
  \href{https://doi.org/10.1103/PhysRevD.88.085038}{\emph{Phys. Rev. D}
  {\bfseries 88} (2013) 085038}
  [\href{https://arxiv.org/abs/1307.7696}{{\ttfamily 1307.7696}}].

\bibitem{Kallosh:2013yoa}
R.~Kallosh, A.~Linde and D.~Roest, \emph{{Superconformal Inflationary
  $\alpha$-Attractors}},
  \href{https://doi.org/10.1007/JHEP11(2013)198}{\emph{JHEP} {\bfseries 11}
  (2013) 198} [\href{https://arxiv.org/abs/1311.0472}{{\ttfamily 1311.0472}}].

\bibitem{Carrasco:2015pla}
J.J.M.~Carrasco, R.~Kallosh and A.~Linde, \emph{{$\alpha $-Attractors: Planck,
  LHC and Dark Energy}},
  \href{https://doi.org/10.1007/JHEP10(2015)147}{\emph{JHEP} {\bfseries 10}
  (2015) 147} [\href{https://arxiv.org/abs/1506.01708}{{\ttfamily
  1506.01708}}].

\bibitem{Galante:2014ifa}
M.~Galante, R.~Kallosh, A.~Linde and D.~Roest, \emph{{Unity of Cosmological
  Inflation Attractors}},
  \href{https://doi.org/10.1103/PhysRevLett.114.141302}{\emph{Phys. Rev. Lett.}
  {\bfseries 114} (2015) 141302}
  [\href{https://arxiv.org/abs/1412.3797}{{\ttfamily 1412.3797}}].

\bibitem{Broy:2015qna}
B.J.~Broy, M.~Galante, D.~Roest and A.~Westphal, \emph{{Pole inflation
  \textemdash{} Shift symmetry and universal corrections}},
  \href{https://doi.org/10.1007/JHEP12(2015)149}{\emph{JHEP} {\bfseries 12}
  (2015) 149} [\href{https://arxiv.org/abs/1507.02277}{{\ttfamily
  1507.02277}}].

\bibitem{Terada:2016nqg}
T.~Terada, \emph{{Generalized Pole Inflation: Hilltop, Natural, and Chaotic
  Inflationary Attractors}},
  \href{https://doi.org/10.1016/j.physletb.2016.07.058}{\emph{Phys. Lett. B}
  {\bfseries 760} (2016) 674}
  [\href{https://arxiv.org/abs/1602.07867}{{\ttfamily 1602.07867}}].

\bibitem{Saikawa:2017wkg}
K.~Saikawa, M.~Yamaguchi, Y.~Yamashita and D.~Yoshida, \emph{{Pole inflation in
  Jordan frame supergravity}},
  \href{https://doi.org/10.1088/1475-7516/2018/01/031}{\emph{JCAP} {\bfseries
  01} (2018) 031} [\href{https://arxiv.org/abs/1709.03440}{{\ttfamily
  1709.03440}}].

\bibitem{Tsamis:2003px}
N.C.~Tsamis and R.P.~Woodard, \emph{{Improved estimates of cosmological
  perturbations}},
  \href{https://doi.org/10.1103/PhysRevD.69.084005}{\emph{Phys. Rev. D}
  {\bfseries 69} (2004) 084005}
  [\href{https://arxiv.org/abs/astro-ph/0307463}{{\ttfamily
  astro-ph/0307463}}].

\bibitem{Kinney:2005vj}
W.H.~Kinney, \emph{{Horizon crossing and inflation with large eta}},
  \href{https://doi.org/10.1103/PhysRevD.72.023515}{\emph{Phys. Rev. D}
  {\bfseries 72} (2005) 023515}
  [\href{https://arxiv.org/abs/gr-qc/0503017}{{\ttfamily gr-qc/0503017}}].

\bibitem{Cai:2016ngx}
Y.-F.~Cai, J.-O.~Gong, D.-G.~Wang and Z.~Wang, \emph{{Features from the
  non-attractor beginning of inflation}},
  \href{https://doi.org/10.1088/1475-7516/2016/10/017}{\emph{JCAP} {\bfseries
  10} (2016) 017} [\href{https://arxiv.org/abs/1607.07872}{{\ttfamily
  1607.07872}}].

\bibitem{Brown:2015iha}
J.~Brown, W.~Cottrell, G.~Shiu and P.~Soler, \emph{{Fencing in the Swampland:
  Quantum Gravity Constraints on Large Field Inflation}},
  \href{https://doi.org/10.1007/JHEP10(2015)023}{\emph{JHEP} {\bfseries 10}
  (2015) 023} [\href{https://arxiv.org/abs/1503.04783}{{\ttfamily
  1503.04783}}].

\bibitem{Montero:2015ofa}
M.~Montero, A.M.~Uranga and I.~Valenzuela, \emph{{Transplanckian axions!?}},
  \href{https://doi.org/10.1007/JHEP08(2015)032}{\emph{JHEP} {\bfseries 08}
  (2015) 032} [\href{https://arxiv.org/abs/1503.03886}{{\ttfamily
  1503.03886}}].

\bibitem{Rudelius:2015xta}
T.~Rudelius, \emph{{Constraints on Axion Inflation from the Weak Gravity
  Conjecture}}, \href{https://doi.org/10.1088/1475-7516/2015/9/020}{\emph{JCAP}
  {\bfseries 09} (2015) 020}
  [\href{https://arxiv.org/abs/1503.00795}{{\ttfamily 1503.00795}}].

\bibitem{vanBeest:2021lhn}
M.~van Beest, J.~Calder\'on-Infante, D.~Mirfendereski and I.~Valenzuela,
  \emph{{Lectures on the Swampland Program in String Compactifications}},
  \href{https://arxiv.org/abs/2102.01111}{{\ttfamily 2102.01111}}.

\bibitem{Ishibashi:1996ps}
A.~Ishibashi, T.~Koike, M.~Siino and S.~Kojima, \emph{{Compact hyperbolic
  universe and singularities}},
  \href{https://doi.org/10.1103/PhysRevD.54.7303}{\emph{Phys. Rev. D}
  {\bfseries 54} (1996) 7303}
  [\href{https://arxiv.org/abs/gr-qc/9605041}{{\ttfamily gr-qc/9605041}}].

\bibitem{Numasawa:2019juw}
T.~Numasawa and D.~Yoshida, \emph{{Global Spacetime Structure of Compactified
  Inflationary Universe}},
  \href{https://doi.org/10.1088/1361-6382/ab38ed}{\emph{Class. Quant. Grav.}
  {\bfseries 36} (2019) 195003}
  [\href{https://arxiv.org/abs/1901.03347}{{\ttfamily 1901.03347}}].

\bibitem{1967rta1.book..160M}
C.W.~{Misner}, \emph{{Taub-Nut Space as a Counterexample to almost anything}},
  in \emph{Relativity Theory and Astrophysics. Vol.1: Relativity and
  Cosmology}, J.~{Ehlers}, ed., vol.~8, p.~160 (1967).

\bibitem{Berkooz:2004re}
M.~Berkooz, B.~Pioline and M.~Rozali, \emph{{Closed strings in Misner space:
  Cosmological production of winding strings}},
  \href{https://doi.org/10.1088/1475-7516/2004/08/004}{\emph{JCAP} {\bfseries
  08} (2004) 004} [\href{https://arxiv.org/abs/hep-th/0405126}{{\ttfamily
  hep-th/0405126}}].

\bibitem{Siegel:1993xq}
W.~Siegel, \emph{{Two vierbein formalism for string inspired axionic gravity}},
  \href{https://doi.org/10.1103/PhysRevD.47.5453}{\emph{Phys. Rev. D}
  {\bfseries 47} (1993) 5453}
  [\href{https://arxiv.org/abs/hep-th/9302036}{{\ttfamily hep-th/9302036}}].

\bibitem{Hull:2009mi}
C.~Hull and B.~Zwiebach, \emph{{Double Field Theory}},
  \href{https://doi.org/10.1088/1126-6708/2009/09/099}{\emph{JHEP} {\bfseries
  09} (2009) 099} [\href{https://arxiv.org/abs/0904.4664}{{\ttfamily
  0904.4664}}].

\bibitem{Hohm:2010jy}
O.~Hohm, C.~Hull and B.~Zwiebach, \emph{{Background independent action for
  double field theory}},
  \href{https://doi.org/10.1007/JHEP07(2010)016}{\emph{JHEP} {\bfseries 07}
  (2010) 016} [\href{https://arxiv.org/abs/1003.5027}{{\ttfamily 1003.5027}}].

\bibitem{Hohm:2010pp}
O.~Hohm, C.~Hull and B.~Zwiebach, \emph{{Generalized metric formulation of
  double field theory}},
  \href{https://doi.org/10.1007/JHEP08(2010)008}{\emph{JHEP} {\bfseries 08}
  (2010) 008} [\href{https://arxiv.org/abs/1006.4823}{{\ttfamily 1006.4823}}].

\bibitem{Zwiebach:2011rg}
B.~Zwiebach, \emph{{Double Field Theory, T-Duality, and Courant Brackets}},
  \href{https://doi.org/10.1007/978-3-642-25947-0_7}{\emph{Lect. Notes Phys.}
  {\bfseries 851} (2012) 265}
  [\href{https://arxiv.org/abs/1109.1782}{{\ttfamily 1109.1782}}].

\bibitem{Hohm:2016lim}
O.~Hohm, U.~Naseer and B.~Zwiebach, \emph{{On the curious spectrum of duality
  invariant higher-derivative gravity}},
  \href{https://doi.org/10.1007/JHEP08(2016)173}{\emph{JHEP} {\bfseries 08}
  (2016) 173} [\href{https://arxiv.org/abs/1607.01784}{{\ttfamily
  1607.01784}}].

\bibitem{Noumi:2020php}
T.~Noumi, K.~Saito, J.~Soda and D.~Yoshida, \emph{{$O(d,d;\mathbb{Z})$
  invariant Fierz-Pauli massive gravity}},
  \href{https://doi.org/10.1103/PhysRevD.103.046011}{\emph{Phys. Rev. D}
  {\bfseries 103} (2021) 046011}
  [\href{https://arxiv.org/abs/2010.10871}{{\ttfamily 2010.10871}}].

\end{thebibliography}\endgroup
\end{document}